\documentclass[onecollarge,natbib]{svjour2}
\bibpunct{[}{]}{;}{n}{}{,} 
\smartqed  
\usepackage{graphicx}
\usepackage{mathptmx}      

\journalname{Few Body Systems}
\begin{document}

\title{Heavy Flavour production as probe of Gluon Sivers Function
}


\author{Rohini M. Godbole, Abhiram Kaushik,Anuradha Misra, Vaibhav Rawoot,  
        Bipin Sonawane 
}


\institute{Rohini M. Godbole  \at
              Indian Institute of Science, Bangalore \\
              \email{rohini@chep.iisc.ernet.in}           
           \and
           \at Abhiram Kaushik \at
              Indian Institute of Science, Bangalore \\
              \email{abhiramb@chep.iisc.ernet.in}           
              \and
           Anuradha Misra  \at
              University of Mumbai, Mumbai \\
              \email{misra@physics.mu.ac.in}           
              \and
           Vaibhav S. Rawoot \at
              University of Mumbai, Mumbai\\
              \email{vaibhavrawoot@gmail.com}
              \and
           Bipin Sonawane \at
              University of Mumbai, Mumbai\\
              \email{bipin.sonawane@physics.mu.ac.in}
}
\date{Received: date / Accepted: date}

\maketitle

\begin{abstract}
Heavy flavour production like $J/\psi$ and D- meson production in scattering of electrons/unpolarized protons off polarized proton target offer promising probes to investigate gluon Sivers function. In this talk, I will summarize  our recent work on trasverse single spin asymmetry in $J/\psi $ -production and D - meson production in  $p p^\uparrow$ scattering using a generalized parton model approach. We compare predictions  obtained using different models of gluon Sivers function within this approach and then, taking  into account the  transverse momentum dependent evolution of the unpolarized parton distribution functions  and gluon Sivers function, we study the effect of evolution on asymmetry.

\end{abstract}
\section{Introduction}
\label{intro}
Gluon Sivers function (GSF) belongs to a class of transverse momentum dependent parton distribution functions, 
collectively called TMDs, which are needed to explain the transverse single spin asymmetries (TSSAs) that 
 arise  in the scattering of a transversely polarized nucleon off an unpolarized nucleon (or virtual photon) 
 target and are due to orbital motion of quarks and gluons or due to recoil of gluons radiated off the active quarks.
TSSA for inclusive process $ A^\uparrow + B \rightarrow C + X $ is defined as
\begin{equation} 
A_N = \frac{d\sigma^\uparrow  -  d\sigma^\downarrow}{d\sigma^\uparrow  + d\sigma^\downarrow} 
\label{an} 
\end{equation} 
where $ d\sigma^\uparrow $ and $ d\sigma^\downarrow $ represent the cross section for scattering of a transversely polarized 
hadron A off an unpolarized hadron (or lepton) B with A being upwards (downwards) polarized with respect to the production plane. The magnitude of these asymmetries observed by HERMES and COMPASS  collaborations~\cite{Airapetian:1999tv,Airapetian:2001eg,Alexakhin:2005iw} was found to be larger than perturbative QCD (pQCD) predictions based on conventional collinear factorization~\cite{D'Alesio:2007jt}. One of the two approaches to explain these large asymmetries is based on a generalized factorization formula called transverse momentum dependent (TMD) factorization, in which  parton distribution functions (PDFs) of collinear factorization theorem are replaced by TMDs, which take into account the spin and intrinsic transverse momentum dependence of PDFs. Gluon Sivers Function (GSF) is one such TMD, which  parametrizes the correlations between the azimuthal distribution of an unpolarized parton and the spin of its transversely polarized parent hadron~\cite{Sivers:1989cc, Sivers:1990fh}:
\begin{equation}
\Delta^Nf_{a/p^\uparrow}(x,k_{\perp a}) \equiv
\hat f_{a/p^\uparrow}(x, {\bf k_{\perp}})-\hat f_{a/p^\downarrow}(x, {\bf k_{\perp}})  
\label{delf1} 
\end{equation}

Quark Sivers function have been studied extensively and their parameterizations have been 
extracted from SIDIS experimental data from HERMES, COMPASS and JLab experiments. 
However, not much information is available about GSF so far.   
Recently, the first  rough estimates  of GSF have been obtained by D'Alesio {\it et al} ~\cite{D'Alesio:2015uta}
by fitting to midrapidity data  on  $p p^\uparrow \rightarrow \pi^0X$ taken by 
PHENIX collaboration at RHIC  and, by  using the quark Sivers parameters extracted earlier from SIDIS data.
One now needs to identify processes for which predictions using these fits can be obtained and compared with data.

 Heavy quark and quarkonium systems are natural probes to study the GSF  as the production is sensitive 
to intrinsic transverse momentum especially at low momentum. SSAs in low virtuality electroproduction of $J/\psi $  have been 
estimated by us using  a generalization of color evaporation model of quarkonium production with TMDPDF's~\cite{Godbole:2012bx, Godbole:2013bca, Godbole:2014tha}. SSA in D Meson production in $p p^\uparrow $ scattering, which can be used as a 
clean probe of Gluon Sivers Function~\cite{Anselmino:2004nk} were calculated using a Generalized 
Parton Model approach with two extreme values of GSF - zero and maximum. It was shown that the gluon contribution dominates over quark contribution at RHIC energy and asymmetries up to 25 $ \% $were predicted with saturated GSF. 

Recent measurement of asymmetry  in $J/\psi$ production at PHENIX  experiment with polarized proton beam suggests 
the need for a phenomenological study of TSSA in  heavy flavor production - both open and closed- as probes of GSF. 
 Here,  I  summarize our recent work on transverse single spin asymmetry in the processes 
$p p^\uparrow  \rightarrow J/\psi + X $ and $p p^\uparrow \rightarrow D + X $, where we  have estimated TSSA using 
various parametrizations of GSF presently available and have also studied the effect of TMD evolution of 
Sivers function on TSSA. 

In Section~\ref{sec:2}, we present different parameterizations of GSF that we have used. In Section~\ref{sec:3}, we have summarized our work in Ref~\cite{Godbole:2016tvq} and in Section~\ref{sec:3}, we present our preliminary results for TSSA in $p p^\uparrow \rightarrow J/\psi + X $. Here, we also compare our estimates with the recent Phenix data. In Section~\ref{sec:4}, we present a summary of our results~\cite{Godbole:2016bv}.

\section{Gluon Sivers Function} 
\label{sec:2}
The standard Gaussian form is used for the unpolarized TMD PDF
\begin{equation}
f_{i/p}(x,k_\perp;Q)=f_{i/p}(x,Q)\frac{1}{\pi\langle k_\perp^2\rangle}e^{-k_\perp^2/\langle k_\perp^2\rangle}\nonumber
\end{equation}
with  $\langle k_\perp^2\rangle=0.25$ {GeV}$^2$ and  $i=q,g$. For the Sivers function, 
we use the parametrization ~\cite{D'Alesio:2015uta}
\begin{equation}
\Delta^N f_{i/p^\uparrow}(x,k_\perp;Q)=2\mathcal{N}_{i}(x)f_{i/p}(x,Q)h(k_\perp)
\frac{e^{-k^2_\perp/\langle k_\perp^2\rangle}}{\pi \langle k_\perp^2\rangle}\nonumber
\end{equation}
with 
\begin{equation}
\mathcal{N}_i(x)=N_i x^{\alpha_i}(1-x)^{\beta_i}\frac{(\alpha_i+\beta_i)^{\alpha_i+\beta_i}}{\alpha_i^{\alpha_i} \beta_i^{\beta_i}} 
\,{,}\,\,\,\,\,\,\,\,\, 
h(k_\perp)=\sqrt{2e}\frac{k_\perp}{M_1}e^{-k_\perp^2/M_1^2}
\end{equation}
and
\begin{equation}
 h(k_\perp) \frac{e^{-k^2_\perp/\langle k_\perp^2\rangle}}{\pi \langle k_\perp^2\rangle}
= \frac{\sqrt{2e}}{\pi} \sqrt\frac{1 - \rho}{\rho} {k_\perp} 
\frac{e^{- k_{\perp}^2 / \rho \langle  k_{\perp}^2 \rangle}}{{\langle  k_{\perp}^2 \rangle}^{3/2}},
\end{equation}
where $N_i$, $\alpha_i$, $\beta_i$, $M_1$ and $\rho$ are the best fit parameter. 

We estimate asymmetries using the recently fitted GSF parameters given in Table \ref{SIDIS-gluon-fits}. 
\begin{table}[t]
\caption{SIDIS parameters for GSF and best fit parameters for $u$ and $d$ quark used in BV parametrization.}
\label{SIDIS-gluon-fits}
\centering
\begin{tabular}{|l|l|l|l|l|l|l|}
\hline
DMP-SIDIS 1 & \multicolumn{2}{l|}{$N_g=0.65$} & $\alpha_g=2.8$ & $\beta_g=2.8$ & $\rho=0.687$ & 
{$\langle k^2_\perp\rangle=0.25$ GeV$^2$} 
\\ \cline{1-6}
DMP-SIDIS 2 & \multicolumn{2}{l|}{$N_g=0.05$} & $\alpha_g=0.8$ & $\beta_g=1.4$ & $\rho=0.576$ &                                                        \\ \cline{1-7}
\end{tabular}
\begin{tabular}{|l|l|l|l|l|l|l|}
\hline
$u$ quark & \multicolumn{2}{l|}{$N_u=0.4$} & $\alpha_u=0.35$ & $\beta_u=0.26$  & 
{$\langle k^2_\perp\rangle=0.25$ GeV$^2$} & { $M_1^2=0.19$ } 
\\ \cline{1-5}
$d$ quark & \multicolumn{2}{l|}{$N_d=-0.97$} & $\alpha_d=0.44$ & $\beta_d=0.90$  &    &                                                    \\ \cline{1-7}
\end{tabular} 
\end{table}
These fits have been obtained by  fitting $pp^\uparrow\to \pi^0+X$ asymmetry data at RHIC~\cite{Adare:2013ekj}. 
We present estimates for SSA in D meson production and $J/\psi$  production  using  these parameters, which we will call DMP fits following Ref.~\cite{Godbole:2016tvq}. In case of $J/\psi$ production, we also compare our predictions with those obtained using  parameters we have used in our earlier work, which we call BV parameters as those  were fitted using a model given by Boer and Vogelsang~\cite{Boer:2003tx}. This parametrization is expressed as
\begin{equation} \mathcal{N}_g(x) = \frac{ \mathcal{N}_u + \mathcal{N}_d } {2} \,\,\,\,\,\,\,\, {(BV-a)}
\,\,\,\,\, and \,\,\,\,\,
 \mathcal{N}_g(x) = \mathcal{N}_d \,\,\,\,\,\,\,\,\,\, {(BV-b)}. \end{equation}

The difference between DMP-SIDIS1 and DMP-SIDIS2 parameters is that the former were fitted using a parametrization of quark Sivers function which took into account only $u$ and $d$ quark flavors while for the latter, $s$ quark contribution was also included. The best fits parameter for up and down valance quark Sivers function we have used are given in Ref.~\cite{Anselmino:2011gs}. We also study the effect of QCD evolution of TMD PDF's using the formalism of Ref.~\cite{Collins:1984kg}. For this study we use the BV parameters only  for  both DGLAP and TMD evolution cases, since the DMP fits  took into account  DGLAP evolution only. 

\section{Transverse Single Spin Asymmetry in $p p^\uparrow \rightarrow DX$}
\label{sec:3}
We estimate SSA in D Meson production in $p p^\uparrow $ scattering using the Generalized Parton Model approach. 
The numerator and denominator of asymmetry are given by~\cite{Anselmino:2004nk},

\begin{eqnarray}
\frac{E_D \, d\sigma^{p^\uparrow  p \to DX}} {d^{3} {\bf p_D}} -
\frac{E_D \, d\sigma^{p^\downarrow p \to DX}} {d^{3} {\bf p_D}} 
&& = \>
\int dx_a \, dx_b  \, dz \, d^2 \mathbf{k}_{\perp a} \, d^2 \mathbf{k}_{\perp b} \, 
d^3 \mathbf{k}_{D} \, 
\delta (\mathbf{k}_{D} \cdot \hat{\bf p}_c) \, 
\delta (\hat s +\hat t +\hat u - 2m_c^2) \> 
{\mathcal C}(x_a,x_b,z,\mathbf{k}_D) \nonumber \\
&& 
  \Biggl\{ \sum_q
\left[\Delta^N f_{q/p^{\uparrow}}(x_a,\mathbf{k}_{\perp a}) \>  f_{\bar q/p}(x_b, 
\mathbf{k}_{\perp b}) \>
\frac{d \hat{\sigma}^{q \bar q \to c \bar c}}
{d\hat t}(x_a, x_b, \mathbf{k}_{\perp a}, \mathbf{k}_{\perp b}, \mathbf{k}_D) \>
 D_{D/c}(z,\mathbf{k}_D)
 \right]  \nonumber \\
&&  +
\left[ \Delta ^N f_{g/p^{\uparrow}}(x_a,\mathbf{k}_{\perp a}) \>  f_{g/p}(x_b, 
\mathbf{k}_{\perp b}) \>
\frac{d \hat{\sigma}^{gg \to c \bar c}}
{d\hat t}(x_a, x_b, \mathbf{k}_{\perp a}, \mathbf{k}_{\perp b}, \mathbf{k}_D) \>
 D_{D/c}(z,\mathbf{k}_D) \right] \Biggr\} \>\nonumber 
\end{eqnarray}
\begin{eqnarray}
\frac{E_D \, d\sigma^{p^\uparrow p \to DX}} {d^{3} {\bf p_D}} +
\frac{E_D \, d\sigma^{p^\downarrow p \to DX}} {d^{3} {\bf p_D}}
\nonumber
&& = 2
\int dx_a \, dx_b \, dz\, d^2 \mathbf{k}_{\perp a} \, d^2 \mathbf{k}_{\perp b} \, 
d^3 \mathbf{k}_D \, 
\delta (\mathbf{k}_D \cdot \hat{\bf p}_c) \, 
\delta (\hat s +\hat t +\hat u - 2m_c^2) \, {\mathcal C}(x_a,x_b,z,\mathbf{k}_D)
\nonumber \\
&&  \times \, \Biggl\{ \sum_q
\left[ \hat f_{q/p}(x_a,\mathbf{k}_{\perp a}) \> 
\hat f_{\bar q/p}(x_b, \mathbf{k}_{\perp b}) \>
\frac{d \hat{\sigma}^{q \bar q \to c \bar c}}
{d\hat t}(x_a, x_b, \mathbf{k}_{\perp a}, \mathbf{k}_{\perp b},  \mathbf{k}_D) \>
\hat D_{D/c}(z,\mathbf{k}_D) \right] \nonumber \\
&&  +
\left[ \hat f_{g/p}(x_a,\mathbf{k}_{\perp a}) \> \hat f_{g/p}(x_b, \mathbf{k}_{\perp b}) \>
\frac{d \hat{\sigma}^{gg \to c \bar c}}
{d\hat t}(x_a, x_b, \mathbf{k}_{\perp a}, \mathbf{k}_{\perp b}, \mathbf{k}_D) \>
\hat D_{D/c}(z,\mathbf{k}_D) \right] \Biggr\} \> 
\end{eqnarray}
respectively, where all the symbols have usual meaning~\cite{Godbole:2016tvq}. 

\begin{figure}
\centering
\includegraphics[width=0.37\linewidth]{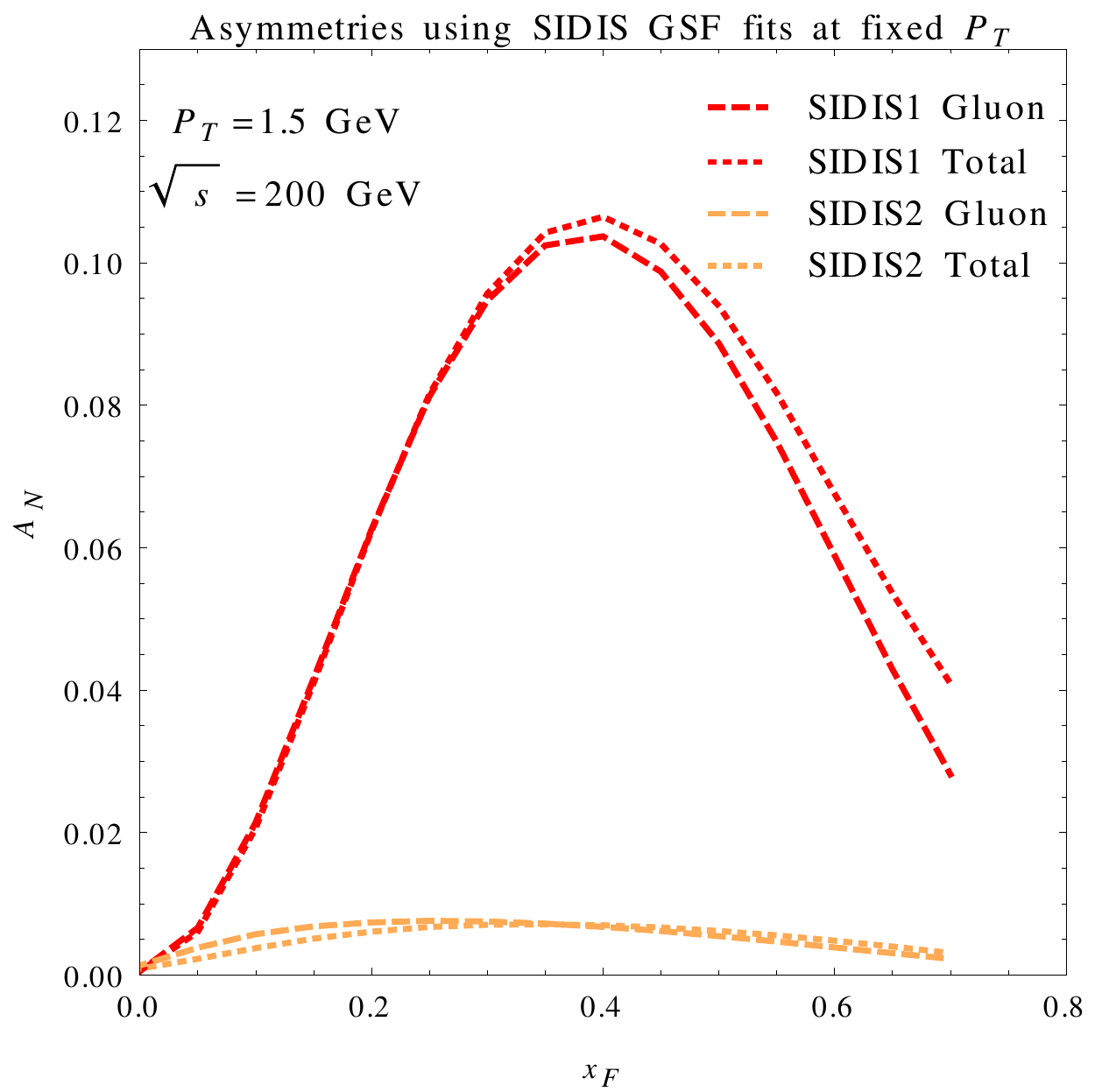}
\includegraphics[width = 0.37\linewidth]{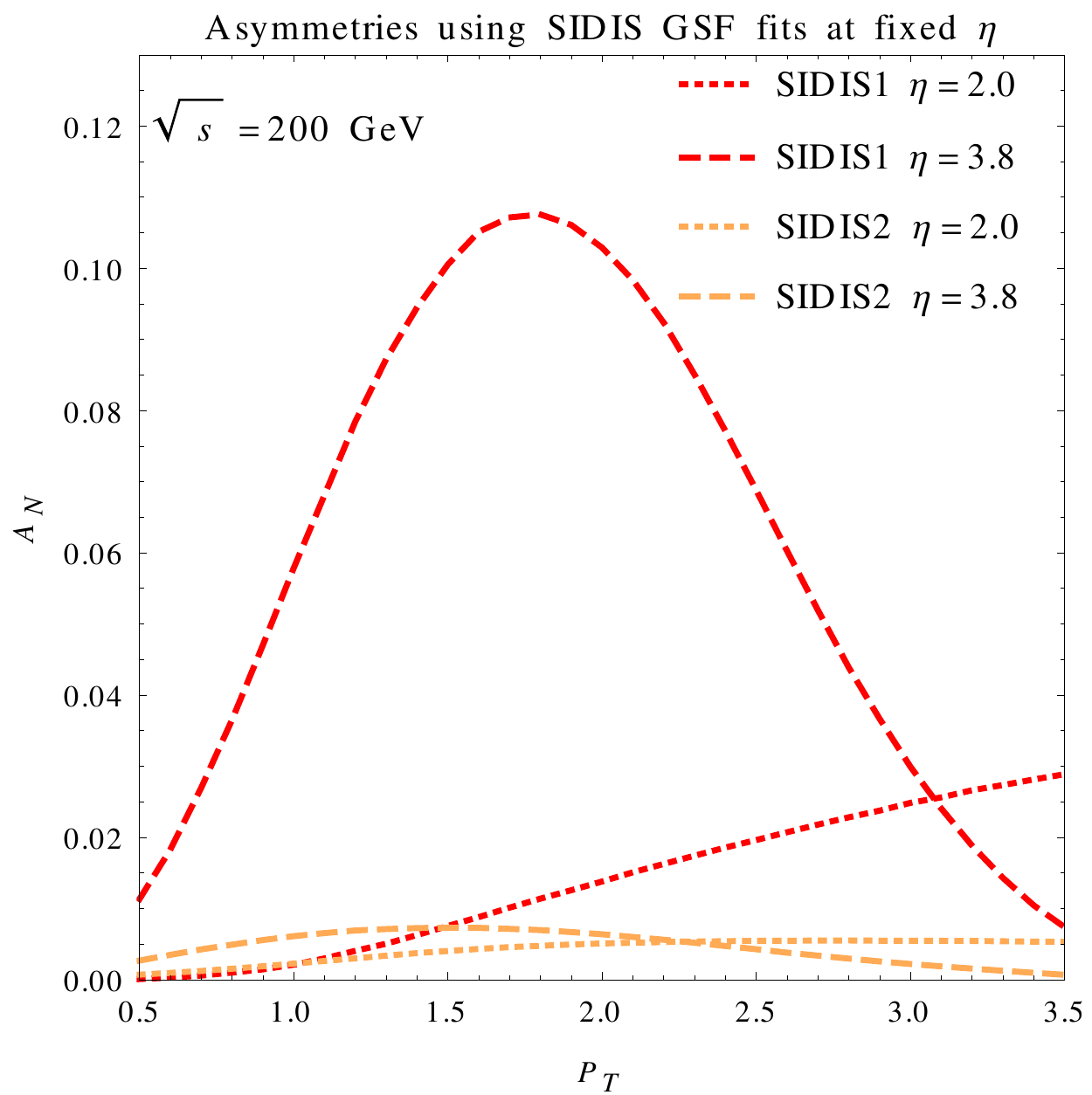}\\
\caption{Asymmetries in D meson production using DMP fits (a)$x_F$ -distribution and (b) $p_T$ -distribution using DMP SIDIS1 and DMP SIDIS 2 fits~\cite{Godbole:2016tvq}}
\label{fig:1}       
\end{figure}
\begin{figure}
\centering
\includegraphics[width=0.37\linewidth]{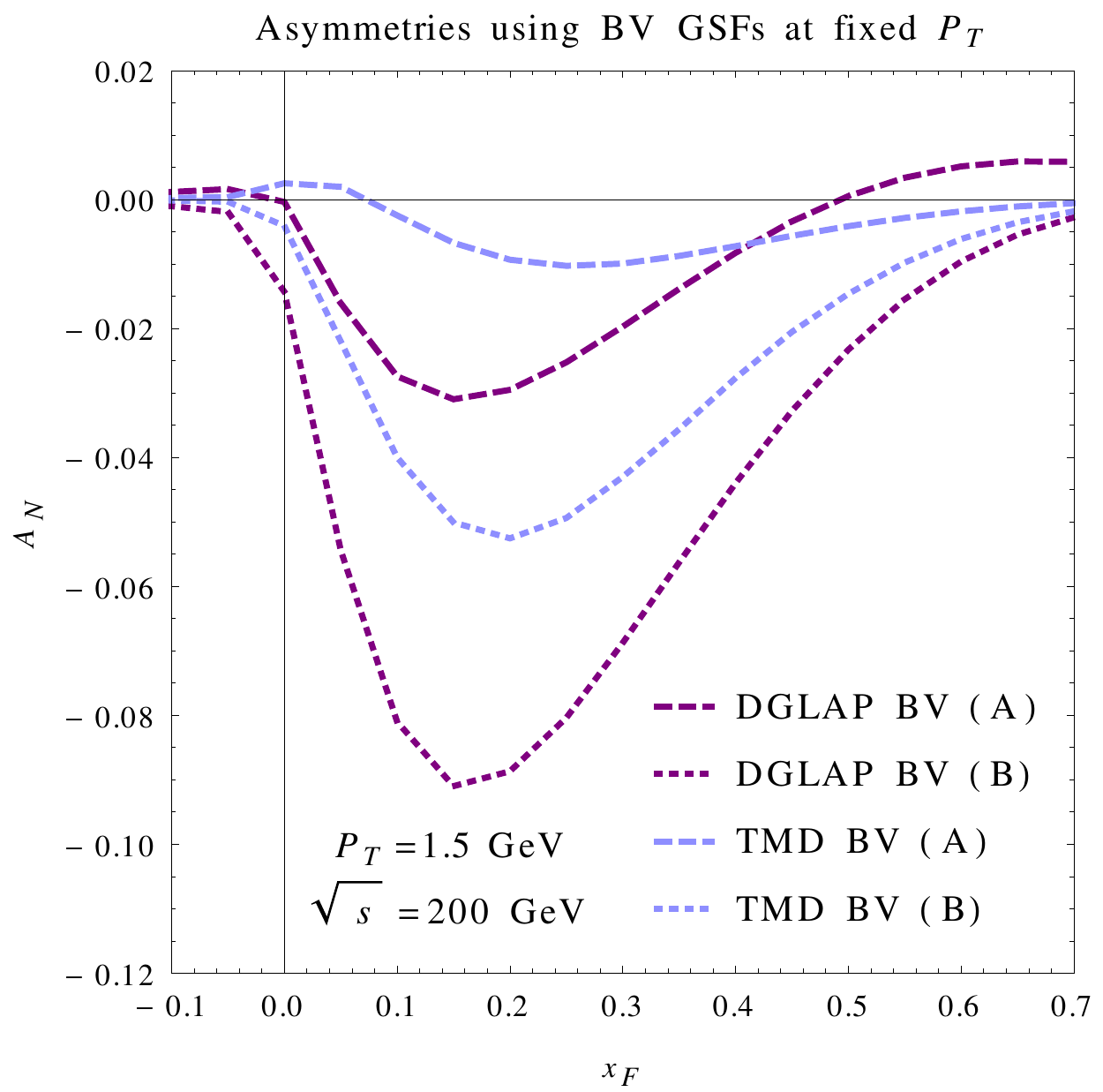}
\includegraphics[width=0.37\linewidth]{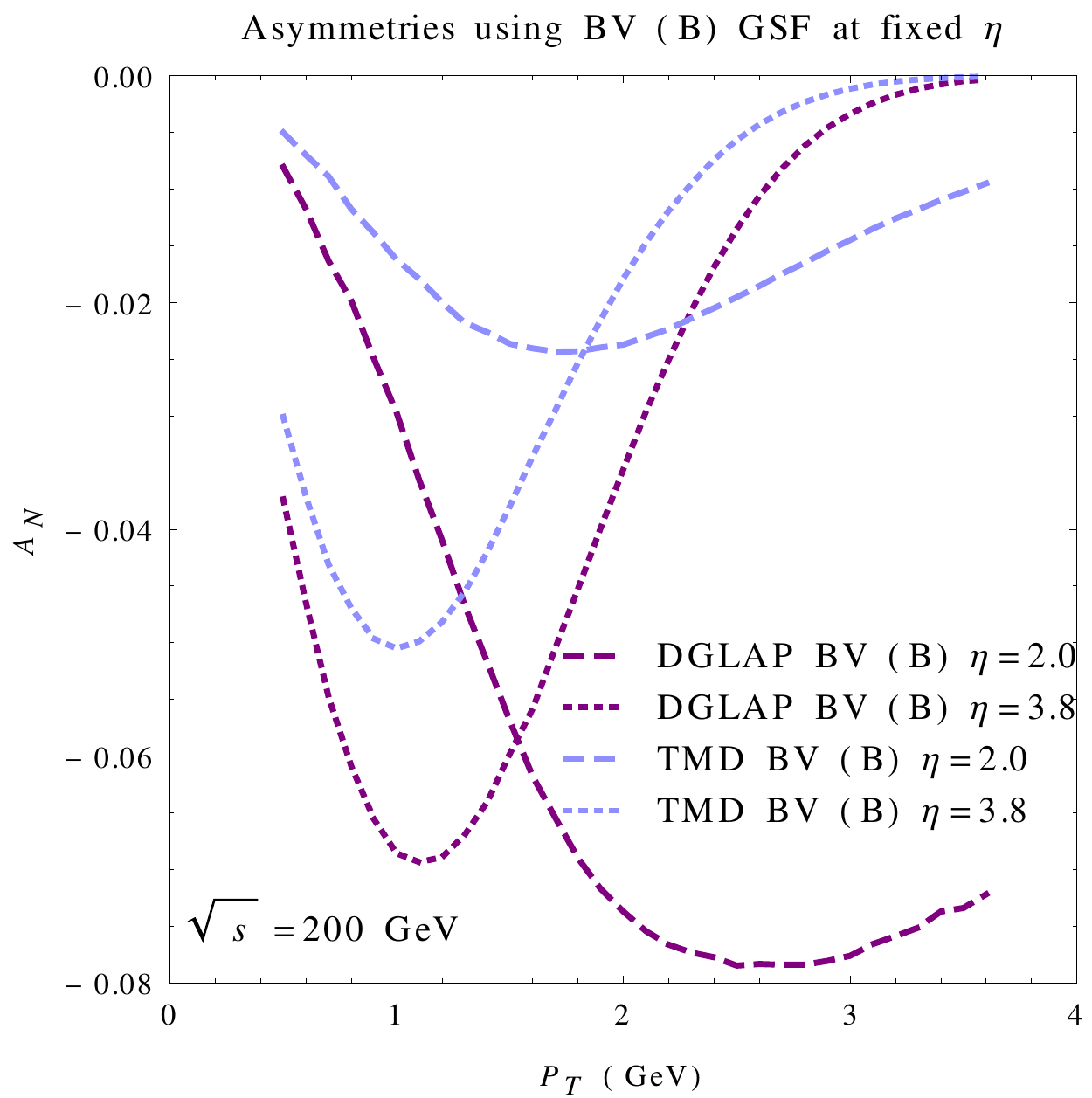}\\
\caption{Comparison of asymmetries in D meson production using BV parameters for GSF fitted using DGLAP evolution and TMD evolution (a)$x_F$ -distribution and (b) $p_T$ -distribution~\cite{Godbole:2016tvq}}
\label{fig:2}       
\end{figure}

We present our estimates of asymmetry  in D meson production at $\sqrt{s} = 200 $ GeV using DMP fits given in Table \ref{SIDIS-gluon-fits}. Estimates of asymmetry at $\sqrt{s} = 115$ GeV and $\sqrt{s} = 500$ GeV can be found in Ref.~\cite{Godbole:2016tvq}. 
In Figure \ref{fig:2}, we have presented comparison of predictions using DGLAP evolved and TMD evolved GSF using BV parameters~\cite{Boer:2003tx}.

\section{Transverse Single Spin Asymmetry in $p p^\uparrow \rightarrow J/\psi X$}
\label{sec:4}
In this section, we present our estimates of TSSA in the process  $p+p^\uparrow \rightarrow J/\psi + X$ 
obtained using a generalized parton model approach and the color evaporation model (CEM) of  $J/\psi$ production~\cite{Godbole:2016bv}:
\begin{eqnarray}
\sigma^{p+p \to J/\psi + X} = F_{J/\psi} \int_{4m^2_c}^{4m^2_D} dM^2_{c\bar c}
\int d x_a d^2 k_{\perp_1} d x_b d^2 k_{\perp_b}
 f_{g/p^\uparrow}(x_a, k_{\perp_a}) f_{g/p}(x_b, k_{\perp_b}) 
\frac{d\hat\sigma^{gg\to c\bar c}}{dM^2_{c\bar c}}
\end{eqnarray}

\begin{figure}[h]
\begin{center}
\includegraphics[width=0.37\linewidth,angle=0]{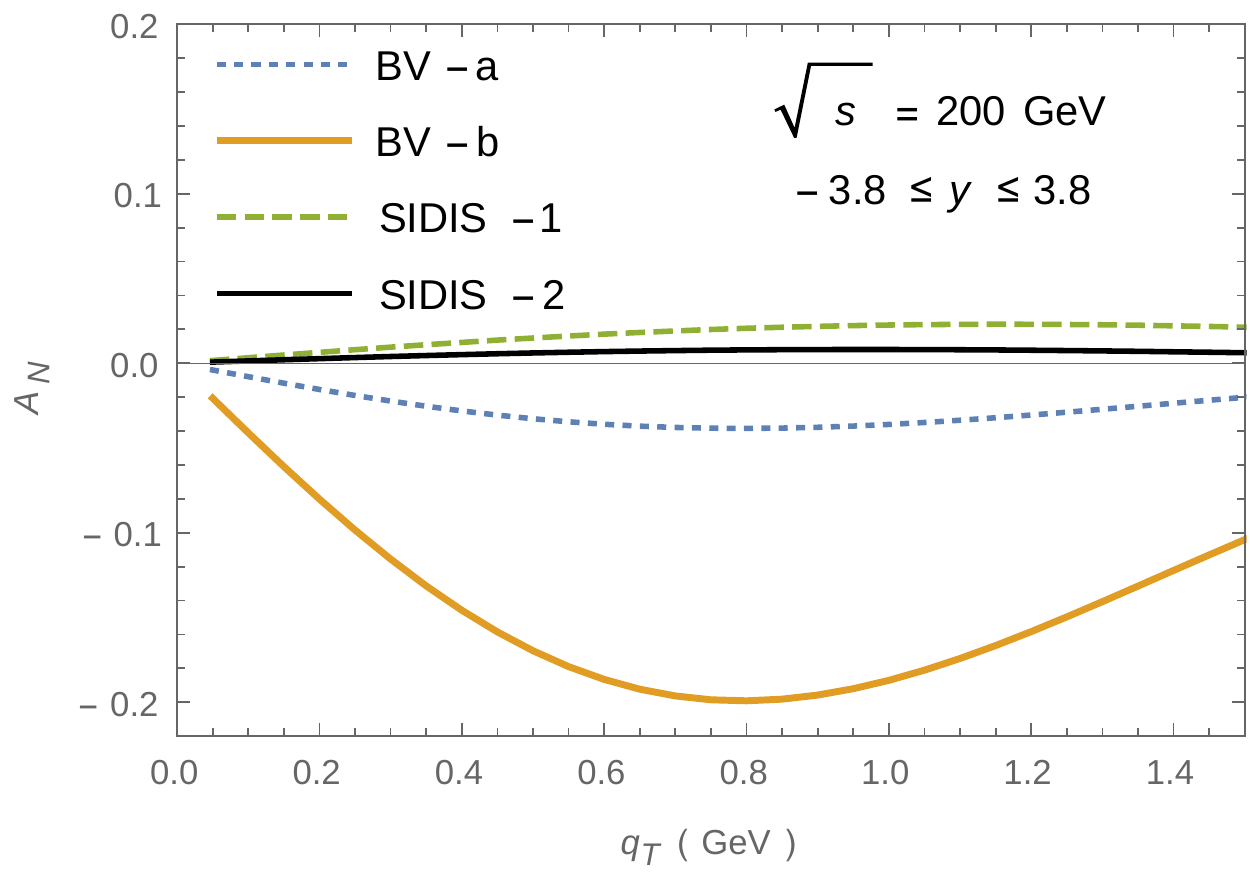}
\includegraphics[width=0.37\linewidth,angle=0]{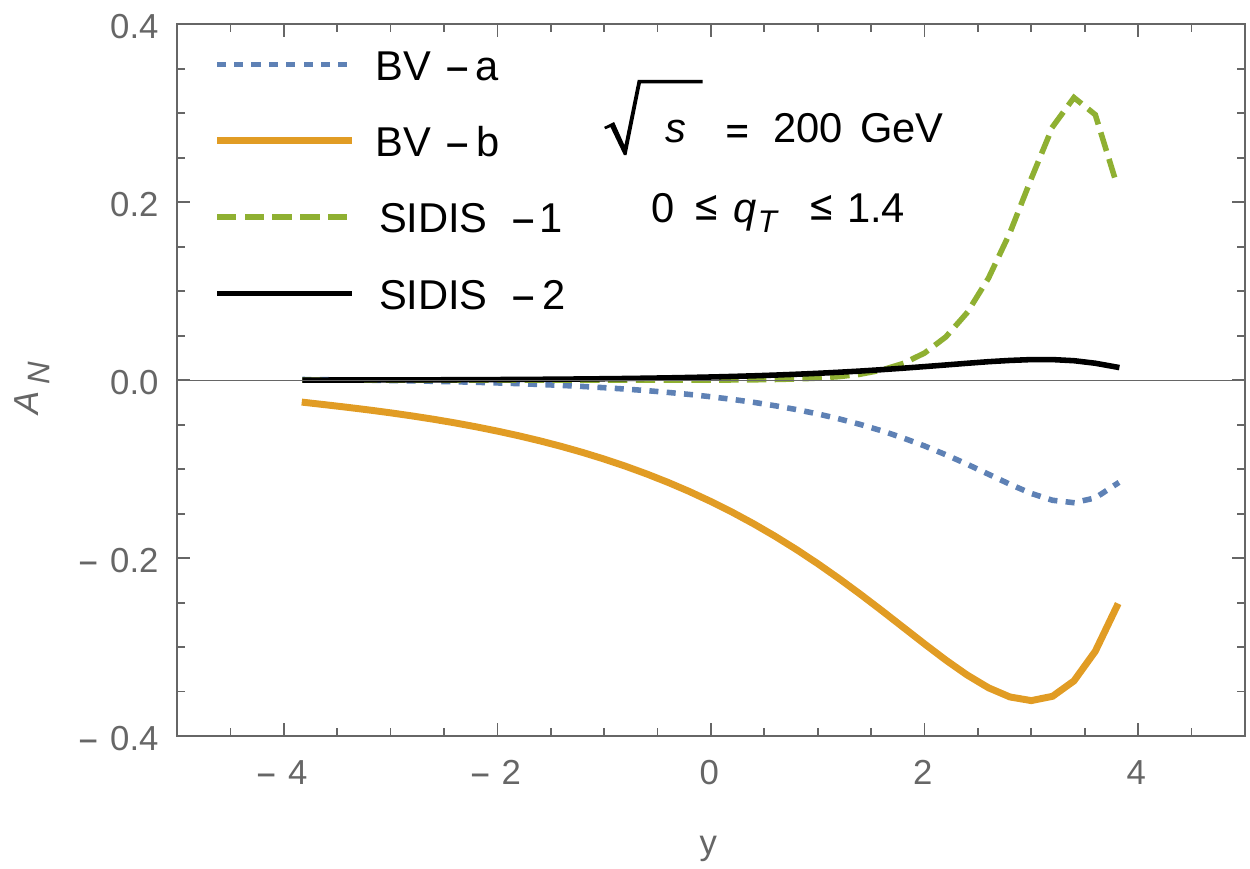}\\
\caption{Predictions of asymmetry in $p^\uparrow+ p\rightarrow J/\psi +X$ at PHENIX ($\sqrt{s}=$ 200 GeV).
$q_T$ and $y$ integration ranges are 0$<q_T<$ 1.4 GeV and -3.8$<y<$3.8 respectively. Left panel and right panel
are plots of $q_T$ and $y$ distribution of asymmetry respectively~\cite{Godbole:2016bv}.}
\label{fig:3}
\end{center}
\end{figure}

In this generalization of CEM, the numerator of the asymmetry in $p+p^\uparrow \rightarrow J/\psi +X$ is parameterized in terms of the Sivers function
\begin{eqnarray}
\frac{d^4 \sigma^{\uparrow}}{dydM^2d^2 {\mathbf q_T}} 
-\frac{d^4\sigma^\downarrow}{dydM^2d^2{\mathbf q_T}}= &&
\frac{1}{s}\int [d^2\mathbf{k}_{\perp a}d^2\mathbf k_{\perp b}]\nonumber 
\Delta^{N}f_{g/p^{\uparrow}}(x_{a},\mathbf {k}_{\perp a})
f_{g/p}(x_{b},\mathbf {k}_{\perp b}) \\
&& \delta^2 ( \mathbf{k}_{\perp a}+\mathbf{k}_{\perp b}-\mathbf q_T)
\hat\sigma_{0}^{g g\rightarrow c\bar{c}}(M^2)
\label{num-ssa}
\end{eqnarray}
The denominator of the asymmetry is 
\begin{eqnarray} 
\frac{d^{4}\sigma^\uparrow}{dydM^2d^2\mathbf q_T}+\frac{d^4\sigma^\downarrow}{dydM^2d^2\mathbf q_T}=&&
\frac{2}{s}\int [d^2\mathbf {k}_{\perp a}d^{2}\mathbf {k}_{\perp b}]
\bigg[f_{g/p}(x_a,\mathbf{k}_{\perp a})
f_{g/p}(x_{2},\mathbf{k}_{\perp b}) \delta^2(\mathbf{k}_{\perp a}+\mathbf{k}_{\perp b}-\mathbf q_T)
\hat\sigma_{0}^{g g\rightarrow c\bar{c}}(M^2)\nonumber \\
&& +f_{q/p}(x_a,\mathbf{k}_{\perp a})
f_{{\bar q}/p}(x_{b},\mathbf{k}_{\perp b}) 
\delta^2(\mathbf{k}_{\perp a}+\mathbf{k}_{\perp b}-\mathbf q_T)
\hat\sigma_{0}^{q {\bar q}\rightarrow c\bar{c}}(M^2)\bigg]
\label{deno-ssa}
\end{eqnarray} 
where, in Eq.(\ref{num-ssa}), we have neglected the contribution of $q {\bar q}$ subprocess as it is negligible compared to gg subprocess. 
In Figure \ref{fig:3}, we present our estimates of asymmetries  calculated using the DMP SIDIS1 and SIDIS2 fits at  $ \sqrt{s} = 200$ GeV. Estimates of asymmetry at $ \sqrt{s} = 115$ GeV  and $500$ GeV can be found in Ref.~\cite{Godbole:2016bv}. 
To investigate the effect of TMD evolution on asymmetry, we  have also calculated asymmetries using BV parameters and  the best fit parameters of the quark Sivers function (used in our earlier work on SSA  in $e+p^\uparrow  \rightarrow J/\psi +X$ and $p+p^\uparrow  \rightarrow D+X$). 
In Figure 4, we compare the predictions with BV parameters obtained using DGLAP evolved and TMD evolved Sivers functions 
to assess the effect of TMD evolution on asymmetries. 
We compare our estimates with the PHENIX data ~\cite{Adare:2013ekj} in figure~\ref{fig:5}. 
We have given estimates in three rapidity regions to compare with PHENIX experimental data.
The backward and forward rapidity regions corresponding to ranges -2.2$\leq y \leq$ -1.2 and 2.2$\leq y \leq$ 1.2 respectively and the mid rapidity region corresponding to range -0.35$\leq y \leq$ -0.35.

\begin{figure}
\begin{center}
\includegraphics[width=0.37\linewidth,angle=0]{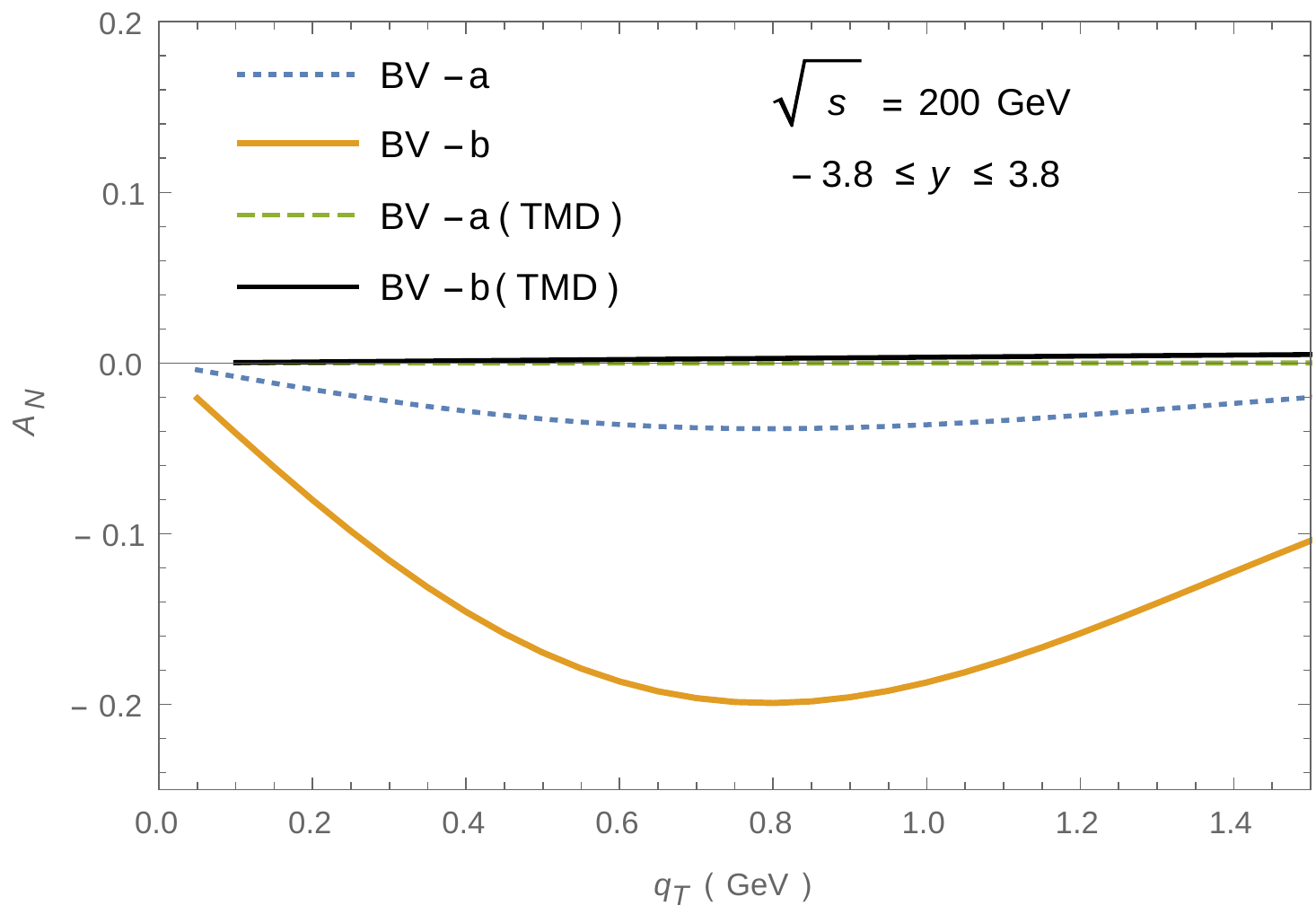}
\includegraphics[width=0.37\linewidth,angle=0]{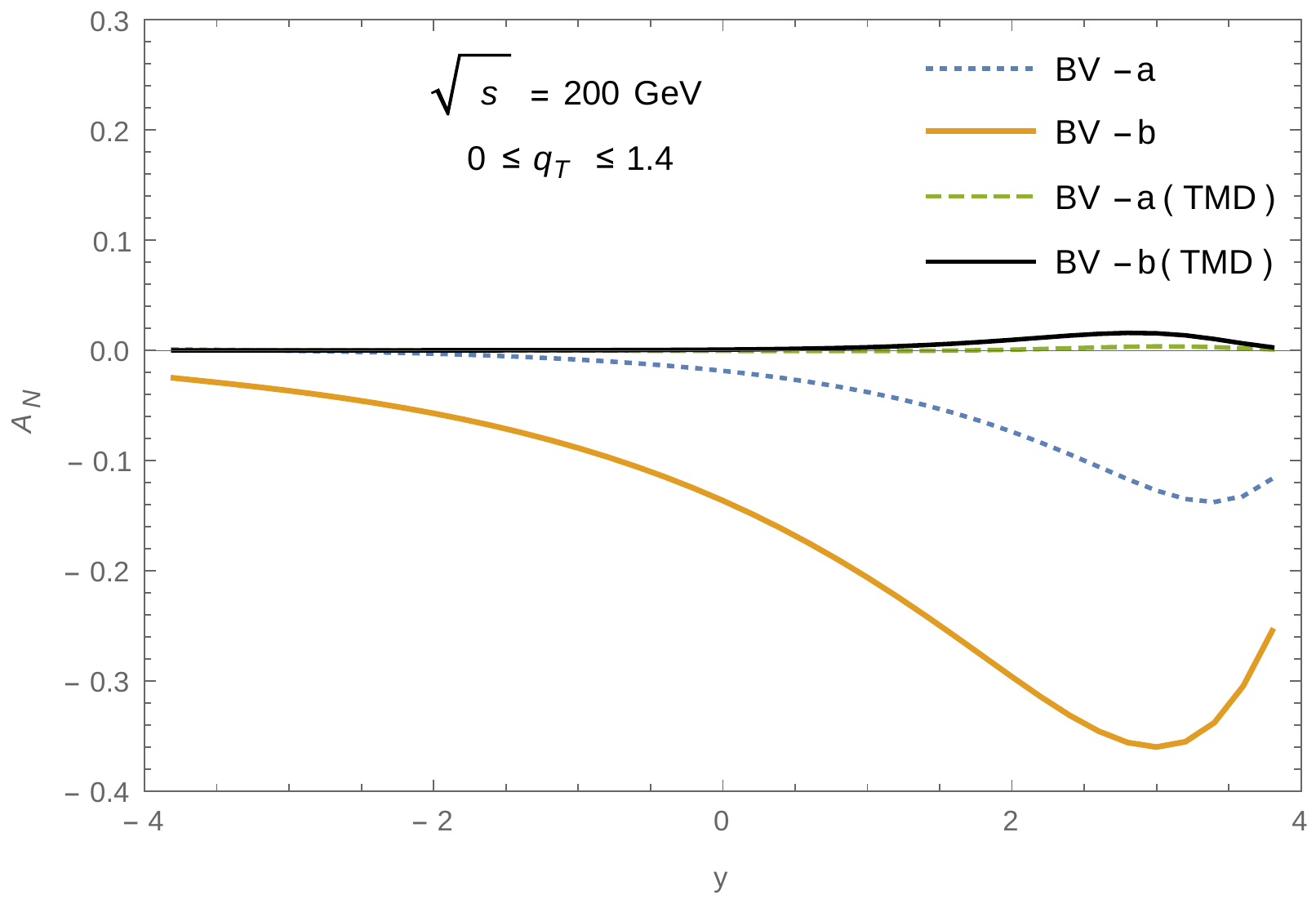}\\
\caption{Predictions of asymmetry with DGLAP and TMD evolution in $p^\uparrow+ p\rightarrow J/\psi +X$ at $\sqrt{s}=$200 GeV.
$q_T$ and $y$ integration ranges are 0$<q_T<$ 1.4 GeV and -3.8$<y<3.8$. Left panel and right panel
    are plots of $q_T$ and $y$ distribution of asymmetry respectively~\cite{Godbole:2016bv}}
\label{fig:4}
\end{center}
\end{figure}

\begin{figure}
\begin{center}
\includegraphics[width=0.37\linewidth,angle=0]{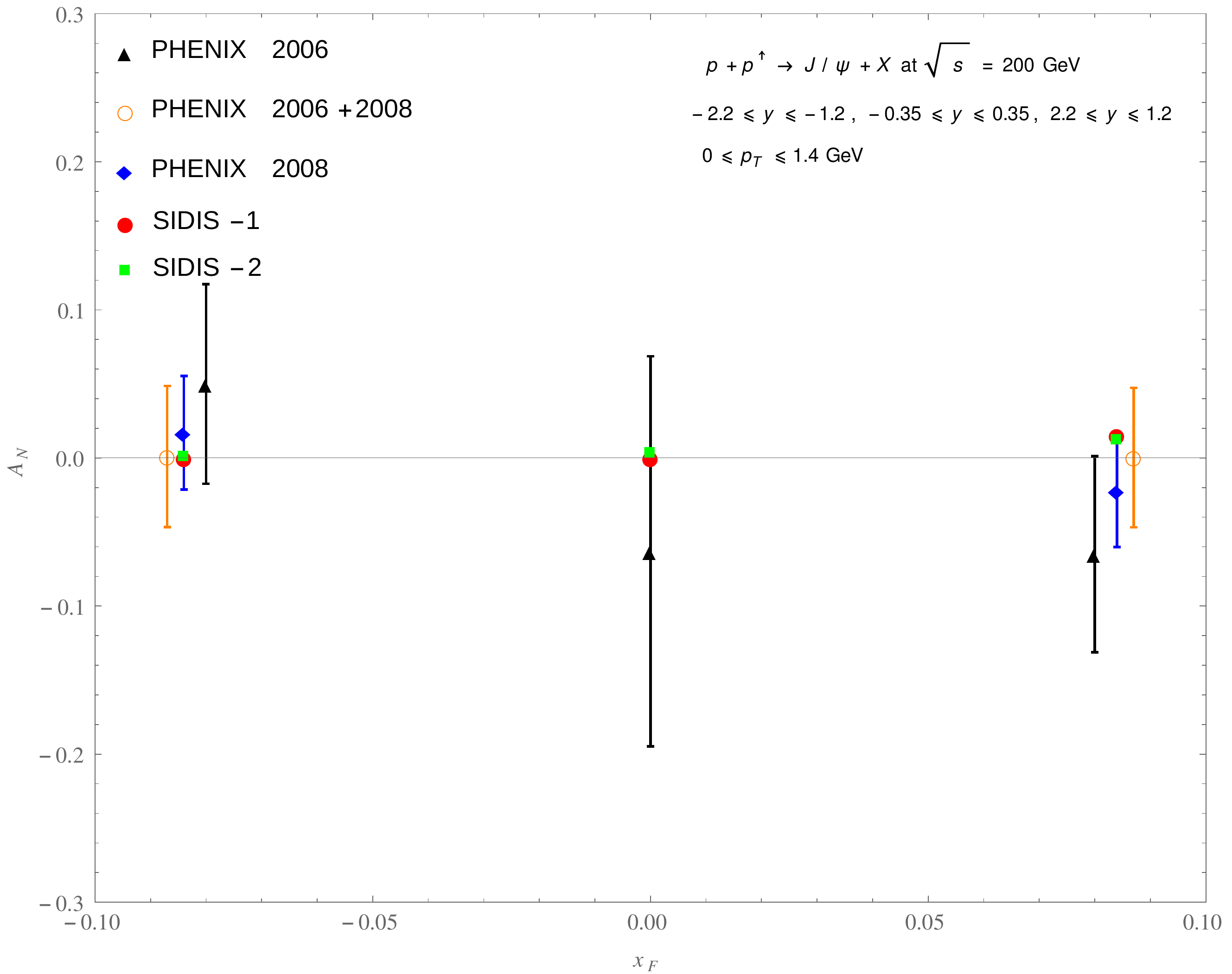}
\includegraphics[width=0.37\linewidth,angle=0]{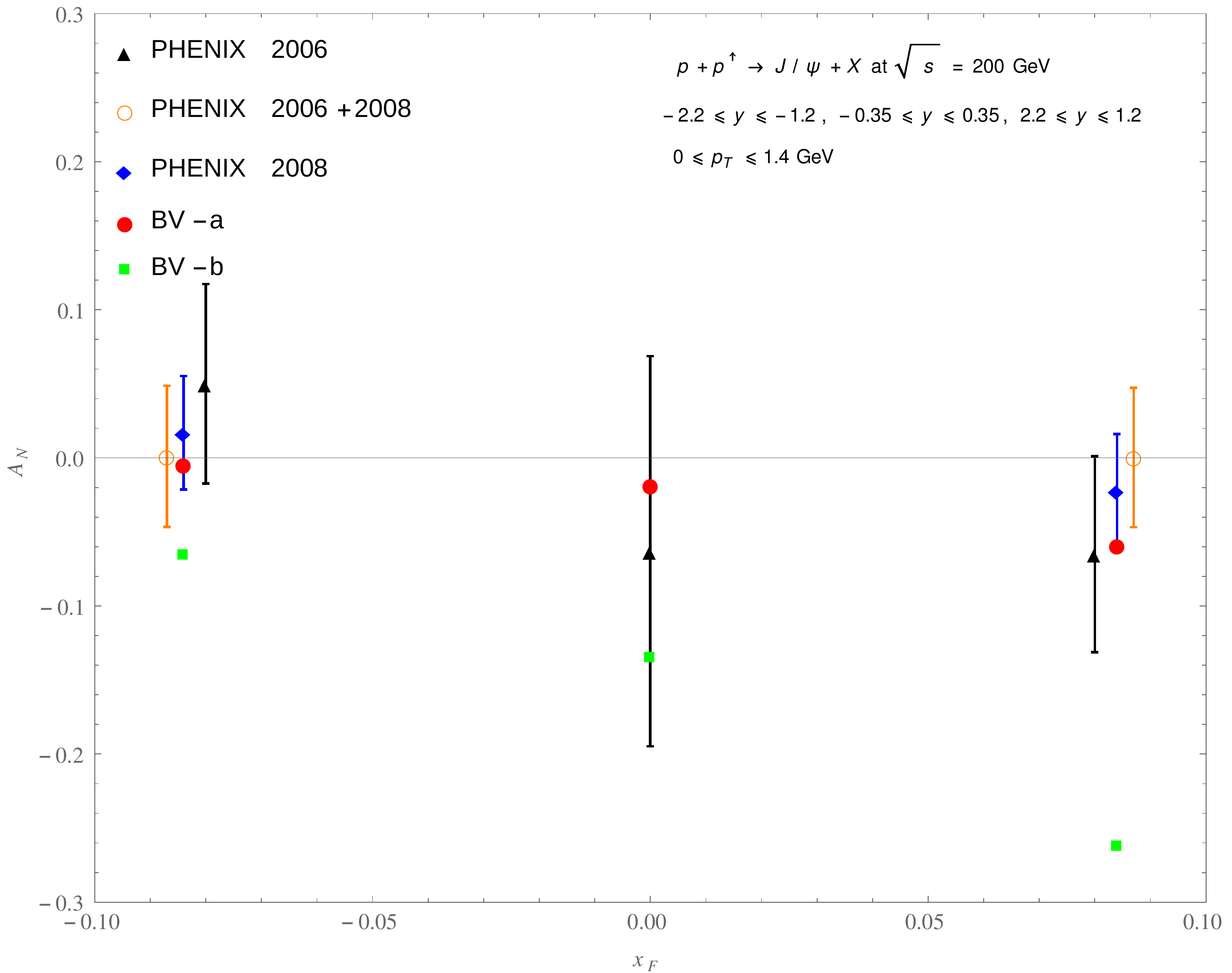}
\caption{Plots of asymmetry in $p^\uparrow+ p\rightarrow J/\psi +X$ at PHENIX experiment ($\sqrt{s}=$ 200 GeV).
 Region of $q_T$ integration is 0$<q_T<1.4$ GeV. Asymmetry is in backward (-2.2$<y<$-1.2), mid (-0.35$<y<$0.35) and forward (1.2$<y<$2.2) rapidity regions.
In left panel we have compared asymmetry obtained using SIDIS-2 and SIDIS-1 parameters with PHENIX data. In right panel we have 
compared asymmetry obtained using BV-a and BV-b parametrization with TMD evolution with PHENIX data~\cite{Godbole:2016bv}}
\label{fig:5}
\end{center}
\end{figure}
\section{Summary}
\label{sec:5}
We have presented estimates of TSSA in D Meson production and $J/\psi$ production in $p p^\uparrow$ scattering at $\sqrt{s} = 200$ GeV. In case of D-Meson production, symmetries are found to be substantial (up to 10\%), while in case of $J/\psi$ production, asymmetries are found to be consistent with the almost zero result of PHENIX experiment. In both cases, the contribution of gluon Sivers function is found to dominate over the quark contribution and the effect of TMD evolution is found to be reduction in the asymmetry estimates. 
Our comparison of estimated asymmetry with PHENIX data shows that asymmetry with SIDIS-2 and SIDIS-1 parameter set is within the experimental uncertainties of PHENIX data. We have also verified  that results obtained using BV-a parameterization is well within experimental uncertainties of PHENIX data, while the estimates of asymmetry with BV-b is not in agreement with PHENIX data~\cite{Godbole:2016bv}. However, after taking into account TMD evolution, both the parameter sets lead to asymmetries that agree with near zero result of experiment. 
\begin{acknowledgements}
A.M and B.S. would like to thank DST, India for financial support under the project  no.  EMR/2014/0000486.  
\end{acknowledgements}
\bibliography{ref.bib}  
\bibliographystyle{unsrt}

\end{document}